\documentclass[12pt]{article}

\usepackage{epsfig}
\usepackage{times}
\usepackage{amsmath}
\usepackage{amssymb}
\usepackage{graphicx}
\usepackage[percent]{overpic}
\usepackage{color}
\usepackage[numbers,sort&compress]{natbib}
\usepackage{subfigure}
\usepackage[T1]{fontenc}
\numberwithin{equation}{section}
\usepackage[
colorlinks=true,
linkcolor=blue,
citecolor=blue,
urlcolor=blue,
anchorcolor=blue,
filecolor=blue,
menucolor=blue
]{hyperref}
\makeatletter
\renewcommand{\tagform@}[1]{\maketag@@@{\textcolor{black}{(\ignorespaces#1\unskip\@@italiccorr)}}}
\renewcommand{\@eqnnum}{\normalfont\normalcolor\textcolor{black}{(\theequation)}}
\renewcommand{\@biblabel}[1]{[#1]}
\long\def\@makecaption#1#2{%
  \vskip\abovecaptionskip
  \sbox\@tempboxa{#1. #2}%
  \ifdim \wd\@tempboxa >\hsize
    #1. #2\par
  \else
    \global \@minipagefalse
    \hb@xt@\hsize{\hfil\box\@tempboxa\hfil}%
  \fi
  \vskip\belowcaptionskip}
\makeatother

\begin{document}

\baselineskip 6mm
\renewcommand{\thefootnote}{\fnsymbol{footnote}}


\newcommand{\nc}{\newcommand}
\newcommand{\rnc}{\renewcommand}




\def\be{\begin{eqnarray}}
\def\ee{\end{eqnarray}}
\def\nn{\nonumber\\}


\def\ct{\cite}
\def\la{\label}
\def\eq#1{(\ref{#1})}


\def\a{\alpha}
\def\b{\beta}
\def\g{\gamma}
\def\G{\Gamma}
\def\d{\delta}
\def\D{\Delta}
\def\e{\epsilon}
\def\et{\eta}
\def\ph{\phi}
\def\Ph{\Phi}
\def\ps{\psi}
\def\Ps{\Psi}
\def\k{\kappa}
\def\l{\lambda}
\def\L{\Lambda}
\def\m{\mu}
\def\n{\nu}
\def\th{\theta}
\def\Th{\Theta}
\def\r{\rho}
\def\s{\sigma}
\def\S{\Sigma}
\def\ta{\tau}
\def\o{\omega}
\def\O{\Omega}
\def\pr{\prime}


\def\half{\frac{1}{2}}

\def\goto{\rightarrow}

\def\na{\nabla}
\def\grad{\nabla}
\def\curl{\nabla\times}
\def\div{\nabla\cdot}
\def\pa{\partial}
\def\fr{\frac}

\def\bra{\left\langle}
\def\ket{\right\rangle}
\def\lb{\left[}
\def\lc{\left\{}
\def\ls{\left(}
\def\lp{\left.}
\def\rp{\right.}
\def\rb{\right]}
\def\rc{\right\}}
\def\rs{\right)}

\def\vac#1{\mid #1 \rangle}


\def\td#1{\tilde{#1}}
\def\check{ \maltese {\bf Check!}}


\def\Tr{{\rm Tr}\,}
\def\det{{\rm det}}
\def\text#1{{\rm #1}}


\def\bc#1{\nnindent {\bf $\bullet$ #1} \\ }
\def\ch {$<Check!>$ }
\def\ss {\vspace{1.5cm}}
\def\inf{\infty}

\begin{titlepage}


\hfill\parbox{3.5cm} 

\vspace{1cm}

\begin{center}
{\Large \bf  Lower-dimensional behavior from factorized infrared geometries along holographic RG flows }

\vskip 1. cm
Chanyong Park$^a$\footnote{e-mail cyong21@gist.ac.kr},

\vskip 1cm

{\it  $^a\,$Department of Physics and Photon Science, Gwangju Institute of Science and Technology \\ 
Gwangju 61005, Korea}\\


\end{center}

\thispagestyle{empty}

\vskip2cm


\centerline{\bf ABSTRACT} \vskip 4mm
We study the emergence of effective lower-dimensional behavior in heavy-scalar two-point functions along holographic RG flows. For charged asymptotically AdS black holes, the equal-time spatial correlator remains exponentially screened, while the temporal correlator exhibits distinct temperature-dependent regimes. Away from extremality, the controlled late-time sector is the Schwarzschild-connected saddle, whose decay rate requires matching to the full geometry. Near extremality, an AdS$_2$ throat supports an additional sector whose leading decay rate is fixed by the throat data. For RN-AdS$_4$, sufficiently near extremality, this throat contribution dominates at asymptotically late times if its matched amplitude is nonzero. At extremality, the thermal throat exponential is replaced by the AdS$_2$ power law. We also study the complete zero-density magnetic-brane flow from AdS$_5$ to AdS$_3\times \mathbb{R}^2$. The transverse heavy-scalar correlator remains exponentially screened, whereas the longitudinal correlator approaches the AdS$_3$/CFT$_2$ power law in the infrared. These results show that factorized infrared geometries can support effective lower-dimensional conformal dynamics whose realization depends on temperature, direction, and the relevant saddle sector, without implying dimensional reduction of the microscopic QFT.

\vspace{2cm}


\end{titlepage}

\renewcommand{\thefootnote}{\arabic{footnote}}
\setcounter{footnote}{0}



\section{Introduction}

The AdS/CFT correspondence provides a geometric description of strongly coupled quantum field theories (QFTs) and their renormalization group (RG) flows \cite{Maldacena:1997re,Gubser:1998bc,Witten:1998qj,Witten:1998zw,Klebanov:1991qa,Henningson:1998gx,Freedman:1998tz,Gubser:1999vj,deBoer:1999tgo,deBoer:2000cz,Skenderis:1999mm,Papadimitriou:2004ap,Heemskerk:2010hk}. Different radial regions of the bulk geometry encode different energy scales of the dual theory. The asymptotic and deep bulk regions describe ultraviolet (UV) and infrared (IR) physics, respectively, so the IR geometry governs the long-distance behavior. Explicit holographic RG flows and nonconformal deformations have been explored in a variety of settings \cite{V65Park2022,V65Kim2017,Park:2022mxj,Park:2024omh}, with related constructions widely applied to QCD-like theories and to systems at finite temperature or density \cite{V65Erlich2005,V65Karch2006,V65Sakai2005a,V65Sakai2005b,V65Herzog2007,V65Kim2007,V65Lee2009,V65Park2010,V65Park2012,V65Park2013,V65Erdmenger2021,V65Park2024MPLA}. Although the microscopic boundary QFT remains defined in its original spacetime dimension, the IR geometry may factorize and thereby support effectively lower-dimensional dynamics. A natural question is then how such lower-dimensional IR structure is manifested in observables of the original higher-dimensional theory.

For a $d$-dimensional boundary QFT, such effective lower-dimensional behavior arises for QFTs dual to extremal charged AdS black holes, whose near-horizon geometry contains an AdS$_2\times\mathbb{R}^{d-1}$ factor and is associated with semi-local critical behavior \cite{Chamblin:1999tk,Iqbal:2011in,Almheiri:2014cka,Hotta:2009bm,Iqbal:2011aj,Faulkner:2009wj,Sachdev:2010um,V65Chamon2011,V65SachdevYe1993,V65MaldacenaStanford2016,V65Chowdhury2022,V65Cvetic2016,V65Gu2020,V65Emparan1999,V65Park2016,Faulkner:2011tm}. Previous studies established temporal AdS$_2$ scaling, finite spatial screening, and the finite-temperature pole structure of near-horizon Green functions \cite{Iqbal:2011in,Iqbal:2011aj,Faulkner:2009wj,Faulkner:2011tm}. The approach to extremality and large-dimension correlators has also been investigated \cite{Arnaudo:2024sen,Ficek:2023qsr,Andrade:2013rra}, while complex geodesic and large-mass Wentzel-Kramers-Brillouin (WKB) methods have related heavy-operator correlators to Lorentzian quasinormal modes and late-time decay \cite{HorowitzHubeny:1999,SonStarinets:2002,HerzogSon:2002,SkenderisVanRees:2008,SkenderisVanRees:2009,Fidkowski:2003nf,Festuccia:2005pi,Festuccia:2008zx,Rodriguez-Gomez:2021pfh}. Lower-dimensional IR behavior also appears in magnetic holographic backgrounds, where magnetic-brane solutions develop an AdS$_3$ factor and exhibit effective two-dimensional infrared dynamics \cite{DHoker:2009mmn,DHoker:2009ixq,DHokerKrausShah:2010,DHokerKraus:2011cm}.

These developments establish the importance of factorized IR geometries, but they do not by themselves imply that every long-distance sector of a boundary correlator is governed solely by the lower-dimensional factor. The full geometry, the direction of separation, and the relevant Lorentzian saddle can remain important. We therefore ask a more specific question. When a lower-dimensional factor emerges in the IR geometry, which kinematic sectors of a neutral heavy-scalar two-point function actually exhibit the associated lower-dimensional conformal scaling and which do not? We study this question in two complementary settings using the same class of neutral heavy-scalar probes.

For charged asymptotically AdS black holes, we first distinguish the complex saddle continuously connected to the Schwarzschild-AdS branch from the additional saddle supported by a near-extremal AdS$_2$ throat. This separation allows us to identify which part of the temporal decay is fixed by local near-horizon data and which part requires matching to the full geometry. At nonzero temperature, the Schwarzschild-connected decay coefficient contains a finite full-geometry matching factor. Near extremality, a parametrically separated AdS$_2$ throat supports an additional sector whose leading decay rate is determined by the throat geometry. The isolated AdS$_2$ Green function contains the same leading large-$\Delta$ thermal damping scale, while the full geometry determines whether the corresponding Lorentzian contribution appears with nonzero residue. At extremality, the finite-temperature throat exponential is replaced by the AdS$_2$ power law, whereas the equal-time spatial correlator remains exponentially screened.

We also consider the complete zero-density magnetic-brane flow from AdS$_5$ to AdS$_3\times\mathbb{R}^2$ \cite{DHoker:2009mmn,DHoker:2009ixq}. Earlier studies established the IR current and stress-tensor sectors of this geometry \cite{DHokerKrausShah:2010}, while scalar correlators have also been studied in related charged magnetic backgrounds \cite{DHokerKraus:2011cm}. Here we probe the complete zero-density flow with the same class of neutral heavy-scalar probes used in the charged-black-hole analysis. The transverse long-distance correlator remains exponentially screened, while the longitudinal correlator approaches the power law governed by the AdS$_3$ factor. The latter therefore realizes an effective CFT$_2$ sector within the infrared dynamics of the microscopic four-dimensional QFT.

These two examples realize lower-dimensional infrared dynamics in different ways. In the charged background, the relevant behavior depends on temperature and on the Lorentzian saddle. In the magnetic background, it depends on the direction in which the neutral heavy-scalar correlator is probed. Together these examples show that a factorized IR geometry can support effective lower-dimensional conformal dynamics without microscopic dimensional reduction and without imposing a universal lower-dimensional scaling law on every kinematic sector of the original theory.

The paper is organized as follows. In Sec. 2, we study spatial and temporal heavy-scalar correlators in charged AdS black holes, including the Schwarzschild-connected, near-extremal AdS$_2$-throat, and exact-extremal regimes. In Sec. 3, we analyze the complete magnetic-brane flow and its longitudinal and transverse correlators. In Sec. 4, we summarize the results.

\section{Two-point functions in the IR limit}

We consider a planar asymptotically AdS$_{d+1}$ black hole with AdS radius $R$. Its holographic dual is a $d$-dimensional quantum field theory at finite temperature and finite charge density. Then, the bulk metric is written as
\be
ds^2 = \fr{R^2}{z^2} \ls - f(z) dt^2 + \fr{1}{f(z)} dz^2  + \d_{ij} dx^i dx^j \rs ,   \la{Metric:general}
\ee
where $i,j=1,\ldots,d-1$. For the class of geometries considered here, we parameterize the blackening factor by
\be
f(z) = \ls 1 - \fr{z^2}{z_h^2} \rs g(z) ,   \la{Ansatz:singleroot}
\ee
with $g(0)=1$ so that the boundary metric has the standard AdS normalization. We assume that $g(z)$ is regular and non-negative outside the horizon. The BTZ blackening factor $f(z)=1-z^2/z_h^2$ is recovered when $g(z)=1$. The Hawking temperature is
\be
T_H = \fr{g(z_h)}{2\pi z_h} . \la{Temperature:general}
\ee
A non-extremal horizon has $g(z_h)>0$. Along the charged families considered here, extremality is reached when $g(z_h)=0$ and $g(z)$ develops a simple zero at $z=z_h$. The blackening factor $f(z)$ then has a double zero at the horizon. The purpose of this section is to determine how the associated infrared geometry is reflected in neutral heavy-scalar two-point functions.

In the heavy-operator limit, a boundary two-point function is determined by the geodesic length $L(t,\vec{x}_1;t,\vec{x}_2)$ connecting two boundary operators \cite{Park:2022mxj,Susskind:1998dq,Solodukhin:1998ec,DHoker:1998vkc,Balasubramanian:1999zv,Louko:2000tp,Kraus:2002iv,Kim:2023fbr,Park:2024pkt, Park:2024omh}
\be
G(\Delta t,\Delta\vec{x}) \equiv \bra O (t_1,\vec{x}_1) \  O (t_2,\vec{x}_2) \ket \sim e^{ - \D L(t_1,\vec{x}_1;t_2,\vec{x}_2) /R }  ,   \la{Proposa:TPF}
\ee
where $\Delta t=t_1-t_2$, $\Delta\vec{x}=\vec{x}_1-\vec{x}_2$, and $\D$ is the UV conformal dimension of the operator dual to a bulk scalar of mass $m$. In the heavy-operator limit, $mR\simeq\D\gg1$. For short spatial and temporal separations, $|t_1-t_2|\ll z_h$ and $|\vec{x}_1-\vec{x}_2|\ll z_h$, the geodesic remains in the asymptotic region and the two-point function reduces to  
\be
G(\Delta t,\Delta\vec{x}) \sim  \fr{1}{ \ls - | \Delta t-i\epsilon | ^2 + |\Delta\vec{x}|^2 \rs^{\D} }  ,      \la{Result: UVCFTco}
\ee
which is the zero-temperature CFT form, with the usual Lorentzian $i\epsilon$ prescription. This is a general feature of thermal correlators since finite thermal corrections are negligible in the UV limit. {In the infrared, the thermal or finite-density state and the bulk geometry modify this behavior} \cite{Kim:2023fbr,Park:2024pkt}. Equation~\eq{Proposa:TPF} is the saddle-point approximation to the heavy bulk propagator and is controlled in this large-$\D$ limit.

\subsection{Spatial correlation function} Since the thermal or finite-density state breaks boost symmetry, temporal and spatial two-point functions need not have the same infrared behavior. We first consider equal-time separation along one spatial direction, denoted by $x$. For boundary points $\lc t,\vec{x}_1\rc$ and $\lc t,\vec{x}_2\rc$, rotational symmetry allows the geodesic to be chosen in the corresponding $x$-$z$ plane. Its length is
\be
L(t,\vec{x}_1;t,\vec{x}_2) =  R \int_{x_1}^{ x_2} d x \ \fr{\sqrt{f +z'^2  }} {z \sqrt{f}} ,
\ee
where the prime denotes a derivative with respect to the chosen boundary coordinate $x$. Translation invariance along this direction gives the conserved quantity
\be
P = - \fr{R \sqrt{f}}{z \sqrt{f + z'^2}}  = - \fr{R}{z_t} ,
\ee
where the last equality is evaluated at a turning point $z_t$ where $z'=0$. 

Introducing a UV cutoff $\e$ near the boundary, the conserved charge allows the boundary separation and geodesic length to be written as functions of the turning point
\be
| \vec{x}_1 - \vec{x}_2 | &=& \int_0^{z_t} dz \fr{2 z}{\sqrt{f} \ \sqrt{z_t^2 - z^2}} , \la{Relation:odistance} \\
L(t,\vec{x}_1;t,\vec{x}_2)  &=&  \int_\e^{z_t} dz  \fr{2 R z_t}{z \, \sqrt{f} \ \sqrt{z_t^2 - z^2} }   \la{Relation:geodesic} ,
\ee
The long-distance IR limit $|\vec{x}_1-\vec{x}_2|\to\infty$ corresponds to $z_t\to z_h$. In this limit, the geodesic length can be reorganized in terms of the boundary separation 
\be
L(t,\vec{x}_1;t,\vec{x}_2)  = \lim_{z_t \to z_h} \ls \fr{R  \, |\vec{x}_1 - \vec{x}_2| }{z_t}+ \fr{2 R}{ z_t} \int_\e^{z_t} dz \fr{\sqrt{z_t^2 -z^2}}{z \sqrt{f}} \rs.   \la{Relation:distgeo}
\ee
The first term grows linearly with the boundary separation, whereas the second contributes only a finite term after the logarithmic UV divergence is subtracted. After renormalization, the leading spatial correlator is 
\be
G(0,\Delta\vec{x}) \sim  e^{-  |\vec{x}_1 - \vec{x}_2|/ \xi } ,  \la{Result:spatialnT}
\ee
{
{where the inverse screening length is $1/\xi$ and the correlation length is given by}
\be
\xi  =  \fr{z_h}{\D}  = \fr{g(z_h)}{2 \pi T_H   \, \D }  .   \la{Result: correlation} 
\ee

{Unlike the UV result in Eq.~\eq{Result: UVCFTco}, the IR correlator is exponentially suppressed by the thermal state and background matter. For fixed horizon data, the inverse screening length is $1/\xi=2\pi T_H \D/g(z_h)$. For a general matter background, $g(z_h)$ is part of the horizon data and must be kept explicit.}
}
 {The zero-temperature limit requires separate treatment. Although Eq.~\eq{Result: correlation} may appear to give a vanishing inverse screening length as $T_H\to0$, $g(z_h)$ vanishes in the same limit.} In the extremal limit, the blackening factor has a double zero
\be
f(z)= \ls 1-\fr{z}{z_h}\rs^2 h(z_h)+\cdots \quad {\rm with}\quad h(z_h)>0 ,
\ee
where $h(z_h)$ is the positive quadratic coefficient of the extremal horizon expansion. In this case, the long-distance limit is again described by $z_t \to z_h$. The boundary distance in Eq.~\eq{Relation:odistance} diverges in this limit. The second term in Eq.~\eq{Relation:distgeo} has the near-horizon form $\sqrt{z_t^2-z^2}/(z_h-z)$ and gives only a finite term after the UV divergence is subtracted. Therefore, the leading geodesic length is $L=R|\vec{x}_1-\vec{x}_2|/z_h$. The exponential correlator in Eq.~\eq{Result:spatialnT} remains valid even at $T_H=0$. {Thus the heavy-operator equal-time correlator retains a finite screening length throughout the temperature range considered. This does not imply a gap in the complete spectrum.}  

\subsection{Temporal correlation function} We next consider Euclidean temporal two-point functions. In the geodesic approximation, the temporal correlator is determined by the Euclidean geodesic length $T(\ta_1,\vec{x};\ta_2,\vec{x})$
\be			 
G(\ta_1-\ta_2,\vec{0}) \sim   e^{- \D T (\ta_1,\vec{x} ; \ta_2 ,\vec{x}) /R}  ,
\ee
with
\be
T (\ta_1,\vec{x}; \ta_2,\vec{x})  = R \int_{\ta_1}^{\ta_2} d \ta \fr{\sqrt{ \dot{z}^2 +  f^2 }}{z \ \sqrt{f}}  ,   \la{Formula: temporalc}
\ee
where the dot denotes a derivative with respect to the Euclidean time $\ta$. Time-translation invariance then gives the conserved charge
\be
H =  - \fr{R \, f^{3/2}}{z \sqrt{\dot{z}^2 + f^2 }}   = - \fr{ R \sqrt{f_t}}{z_t} ,
\ee 
where $f_t=f(z_t)$ denotes the blackening factor at the turning point. From this relation, we can find a relation between $\ta$ and $z$
\be
\fr{dz}{d \ta} = \fr{f \sqrt{  f  z_t^2 - f_t   z^2   }}{z \sqrt{f_t}}  . 
\ee
Using this relation, the boundary time interval and geodesic length can be written as 
\be
| \ta_1 - \ta_2 | &=&  \int_0^{z_t} dz \ \fr{2 z \sqrt{f_t}} {f \sqrt{ f  z_t^2 - f_t   z^2   }}  , \la{Equation:timeinterval1}    \\
T( \ta_1,\vec{x} ;  \ta_2 ,\vec{x}) &=&  \int_\e^{z_t} dz \ \fr{2  R z_t } {z \sqrt{ f  z_t^2 - f_t   z^2   }} .  \la{Equation:geodesic1}
\ee
An exact evaluation of these integrals would determine the temporal two-point function. For $d>2$, however, a nontrivial blackening factor generally prevents a closed analytic result. We therefore focus on the leading infrared behavior.

\subsubsection{Schwarzschild-connected saddle and full-geometry matching}

To study a non-extremal complex saddle that is connected continuously to the Schwarzschild-AdS branch, one must distinguish the local near-horizon normalization from the finite matching to the physical boundary time. In Euclidean signature, the temporal separation itself does not diverge when a real turning point approaches the horizon. The Lorentzian long-time limit is instead obtained after analytic continuation to a complex geodesic saddle. Complexified geodesics, real-time holographic correlators, and their relation to the large-mass WKB approximation and quasinormal frequencies have been studied in Refs.~\cite{HorowitzHubeny:1999,SonStarinets:2002,HerzogSon:2002,SkenderisVanRees:2008,SkenderisVanRees:2009,Fidkowski:2003nf,Festuccia:2005pi,Festuccia:2008zx,Rodriguez-Gomez:2021pfh}. We therefore use the near-horizon geometry only to isolate the local singular structure. All finite information associated with propagation through the outer region is encoded in a matching coefficient that is determined from the full complex saddle.

We introduce the auxiliary blackening factor
\be
\bar{f}(z)=g(z_h)\ls 1-\fr{z^2}{z_h^2}\rs . \la{Ansatz:metric1}
\ee
We write $\bar f_t=\bar f(z_t)$ for its value at the turning point. The auxiliary geometry has the same first-order Rindler zero as the original black hole, but it is not a complete asymptotically AdS geometry. In particular, $\bar f(0)=g(z_h)$ is generally different from one. The time coordinate of the auxiliary geometry cannot in general be identified directly with the physical boundary time of the original geometry. The factor $\sqrt{g(z_h)}$ fixes only the local normalization, while the remaining finite relation depends on the full radial interpolation. We parameterize this information by a dimensionless full-geometry matching coefficient ${\cal M}$ defined through
\be
|\Delta\tau|_{\rm aux}={\cal M}\sqrt{g(z_h)}\,|\tau_1-\tau_2| . \la{Matching:time}
\ee
Here $|\Delta\tau|_{\rm aux}$ denotes the coordinate interval obtained from the auxiliary geodesic integral. The boundary normalization alone would give ${\cal M}=1$. For a general higher-dimensional black hole, ${\cal M}$ is fixed independently by the chosen full complex saddle. In this subsection we follow the branch connected continuously to the Schwarzschild-AdS saddle. The auxiliary geometry therefore isolates the universal local thermal factor, while ${\cal M}$ carries the finite full-geometry information needed for the physical late-time exponent.

Using Eq.~\eq{Equation:timeinterval1} and Eq.~\eq{Equation:geodesic1} with Eq.~\eq{Ansatz:metric1}, we find
\be
|\tau_1-\tau_2|
&=&\fr{1}{{\cal M}\sqrt{g(z_h)}}\int_0^{z_t}dz\,\fr{2z\sqrt{\bar f_t}}{\bar f\sqrt{\bar f z_t^2-\bar f_t z^2}}
=\fr{2z_h}{{\cal M} g(z_h)^{3/2}}\arctan\ls\fr{z_t}{\sqrt{z_h^2-z_t^2}}\rs, \nn
T(\tau_1,\vec{x},\tau_2,\vec{x})
&=&\int_\e^{z_t}dz\,\fr{2Rz_t}{z\sqrt{\bar f z_t^2-\bar f_tz^2}}
=\fr{2R}{\sqrt{g(z_h)}}\log\fr{2z_t}{\e}. \la{Auxiliary:temporal}
\ee
After the Wick rotation $\tau=it$, the turning point is
\be
z_t=iz_h\sinh\ls\fr{{\cal M} g(z_h)^{3/2}|t_1-t_2|}{2z_h}\rs . \la{Auxiliary:turning}
\ee
{This continuation gives the thermal form associated with the local near-horizon structure. It does not determine the full Lorentzian complex saddle.}
At long Lorentzian times, this gives
\be
T(t_1,\vec{x},t_2,\vec{x})\approx2\pi R{\cal M} T_H|t_1-t_2| . \la{Auxiliary:late}
\ee
Consequently, this Schwarzschild-connected finite-temperature saddle has the leading heavy-operator decay envelope
\be
\lim_{|t_1-t_2|\to\infty}|G(t_1-t_2,\vec{0})|
\sim e^{-\Gamma_{\rm SC} |t_1-t_2|} \quad {\rm with}\quad
\Gamma_{\rm SC}\equiv\fr{1}{t_{\rm dec}}=2\pi{\cal M} T_H\D . \la{Result:correlation11}
\ee
The complex saddle can also carry an oscillatory phase. Since the present analysis concerns decay rates, we suppress this phase when quoting the decay envelope in the main text. {The auxiliary geometry only parametrizes the local non-extremal thermal factor in Eq.~\eq{Result:correlation11}. It does not determine the finite coefficient ${\cal M}$ and therefore does not give a geometry-independent decay rate for this full-geometry branch. We determine ${\cal M}$ from the full complex saddle of the original geometry. This is also the large-mass WKB interpretation. Appendix~\ref{Appendix:matching} gives the corresponding complex critical energy and explicit full-saddle matching examples. The Schwarzschild result reproduces the known WKB result of Ref.~\cite{Festuccia:2008zx}.}

For $d=2$, the BTZ geometry provides an exact check. The thermal two-point function is \cite{Kim:2023fbr,Rodriguez-Gomez:2021pfh,Rodriguez-Gomez:2021mkk}
\be
G(\Delta t,\Delta\vec{x})\sim\fr{1}{\left|-\sinh^2\ls\fr{|t_1-t_2|}{2z_h}\rs+\sinh^2\ls\fr{|\vec{x}_1-\vec{x}_2|}{2z_h}\rs\right|^\D} .
\ee
In the convention of Eq.~\eq{Ansatz:singleroot}, BTZ has $g(z_h)=1$ and $T_H=1/(2\pi z_h)$. Since the auxiliary and full blackening factors coincide, ${\cal M}=1$ exactly. The temporal decay constant is therefore
\be
 \Gamma_{\rm SC}=2\pi T_H\D. \la{Result:halftimeBTZ}
\ee
For BTZ, Eq.~\eq{Result: correlation} gives the same numerical scale for spatial screening. This coincidence is special to BTZ. In a general higher-dimensional geometry, the spatial screening length is fixed by the horizon data through Eq.~\eq{Result: correlation}, whereas the temporal decay constant also contains the finite matching coefficient ${\cal M}$ defined by the full complex saddle in Eq.~\eq{Result:correlation11}. The two scales therefore need not coincide. The matching coefficient for a given background is specified only after the full geometry is fixed. The corresponding complex-saddle construction is collected in Appendix~\ref{Appendix:matching}.

\subsubsection{AdS$_2$-throat saddle}

We next isolate the saddle supported by a near-extremal AdS$_2$ throat. This description applies when the blackening factor is close to developing a double zero and a parametrically separated throat region exists. Let $\rho=z_h-z$ denote the radial coordinate distance from the horizon. We denote the positive coefficient of the quadratic term in the near-extremal throat expansion by $h_h$. It approaches the extremal coefficient $h(z_h)$ defined above as $T_H\to0$. The blackening factor then takes the form
\be
f_{\rm throat}(\rho)=4\pi T_H\rho+\fr{h_h}{z_h^2}\rho^2+\cdots
=\fr{h_h}{z_h^2}\rho(\rho+\rho_0)+\cdots \quad {\rm with}\quad \rho_0=\fr{4\pi T_H z_h^2}{h_h}. \la{Throat:generalblackening}
\ee
The corresponding AdS$_2$ radius is $R_2=R/\sqrt{h_h}$. For a neutral scalar of mass $m$ at zero spatial momentum, the exact AdS$_2$ conformal dimension is $\d_{\rm IR}=1/2+\sqrt{1/4+m^2R_2^2}$. The geodesic calculation instead determines the leading large-$\D$ exponent $\D_{\rm IR}=R_2\D/R$. 
Writing $\rho_t=z_h-z_t$ for the turning-point distance, choose a matching point $\rho_m$ inside the throat with $\rho_t,\rho_0\ll\rho_m\ll z_h$. We may set $z\simeq z_t\simeq z_h$ in nonsingular prefactors. The leading throat integrals become
\be
|\Delta\tau|_{\rm throat}&\simeq&2\int_{\rho_t}^{\rho_m}d\rho\,\fr{\sqrt{f_t}}{f_{\rm throat}(\rho)\sqrt{f_{\rm throat}(\rho)-f_t}}
\simeq\fr{1}{\pi T_H}\arctan\ls\fr{\rho_0}{2\sqrt{\rho_t(\rho_t+\rho_0)}}\rs, \nn
T_{\rm throat}&\simeq&\fr{2R}{z_h}\int_{\rho_t}^{\rho_m}\fr{d\rho}{\sqrt{f_{\rm throat}(\rho)-f_t}}
=\fr{2R}{\sqrt{h_h}}\operatorname{arcosh}\ls\fr{2\rho_m+\rho_0}{2\rho_t+\rho_0}\rs . \la{Throat:generalintegrals}
\ee
Here $f_t=f_{\rm throat}(\rho_t)$. Eliminating the turning point and analytically continuing to Lorentzian time gives, up to a time-independent matching normalization,
\be
G_{\rm throat}(t_1-t_2,\vec{0})
\sim\ls\fr{\pi T_H\ell_{\rm IR}}{\sinh(\pi T_H|t_1-t_2|)}\rs^{2\D_{\rm IR}} \quad {\rm with}\quad \D_{\rm IR}=\fr{R_2}{R}\D=\fr{\D}{\sqrt{h_h}}, \la{Throat:thermalcorr}
\ee
where $\ell_{\rm IR}$ is a finite scale fixed by matching the throat to the outer geometry. At $T_H|t_1-t_2|\gg1$, the throat saddle therefore has the decay constant
\be
\Gamma_{\rm throat}=2\pi T_H\D_{\rm IR}
=\fr{2\pi T_H\D}{\sqrt{h_h}} . \la{Throat:decayconstant}
\ee
The throat geometry fixes the leading large-$\D$ damping scale, while the outer region determines the normalization and whether the corresponding Lorentzian contribution appears in the full boundary correlator with nonzero residue. The relevant infrared thermal scale is also encoded in the ingoing finite-temperature AdS$_2$ retarded Green function \cite{SonStarinets:2002,SkenderisVanRees:2009,Faulkner:2009wj,Iqbal:2009fd,Faulkner:2011tm,Ficek:2023qsr}. For the neutral scalar defined above, the lowest pole of the isolated AdS$_2$ Green function has damping $2\pi T_H\d_{\rm IR}$. When the corresponding matched Lorentzian sector is present, $\d_{\rm IR}=\D_{\rm IR}+O(\D^0)$ implies that its leading large-$\D$ damping agrees with the analytically continued geodesic result in Eq.~\eq{Throat:decayconstant}. The full geometry is still required to determine the residue and the competition with the Schwarzschild-connected branch.

Along a family of solutions approaching extremality, a parametrically separated AdS$_2$ radial window exists only sufficiently close to the extremal limit. Away from that regime, the independent throat description is not controlled. When the throat is present, the outer geometry fixes its matched amplitude, while the leading decay constant is given by Eq.~\eq{Throat:decayconstant}. Which saddle controls the complete Lorentzian correlator is a global matching and Stokes problem rather than a sharp local phase boundary.

\subsubsection{Extremal AdS$_2$ saddle}

{The zero-temperature geometry must be treated as a distinct extremal saddle. From Eq.~\eq{Throat:decayconstant}, the AdS$_2$-throat decay constant satisfies $\Gamma_{\rm throat}\to0$ as $T_H\to0$. The exponential suppression of the finite-temperature throat sector therefore disappears. The extremal geometry has an exact AdS$_2$ throat, so we analyze the temporal two-point function in Eq.~\eq{Formula: temporalc} directly at $T_H=0$.

As in the previous finite-temperature case, it is difficult to perform the integrals in Eq.~\eq{Equation:timeinterval1} and Eq.~\eq{Equation:geodesic1} exactly. Therefore, we focus on the leading IR behavior. To do so, we introduce another blackening factor $\bar{F}$, which approximates the near-horizon geometry well in the extremal limit,
\be
\bar{F} = \ls 1 -  \fr{z}{z_h} \rs^2 \,  h(z_h) .
\ee
The coefficient $h(z_h)$ is determined from the extremal blackening factor by 
\be
h(z_h) = \lim_{z \to z_h} \fr{f(z)}{ \ls 1 -  \fr{z}{z_h} \rs^2} .
\ee
At extremality, $h_h\to h(z_h)$ and the AdS$_2$ radius introduced in the near-extremal analysis becomes
\be
R_2=\fr{R}{\sqrt{h(z_h)}} .
\ee
A finite matching rescaling of the time coordinate changes only the overall normalization of the power-law correlator and does not affect the exponent. For convenience, we define a rescaled throat time by
\be
|\bar{\ta}_1 - \bar{\ta}_2 |    \equiv  \sqrt{h(z_h)}  \,  |\ta_1 - \ta_2| .
\ee
}

Defining $\bar F_t=\bar F(z_t)$ at zero temperature with $z_t \to z_h$, the leading boundary time interval and geodesic length then take the form 
\be
| \ta_1 - \ta_2 | &=& 
\fr{1}{\sqrt{h(z_h)}}   \int_0^{z_t} dz \ \fr{2 z \sqrt{\bar{F}_t}} {\bar{F} \sqrt{\bar{F} z_t^2 - \bar{F}_t z^2}} \approx   \fr{2 z_h^2}{ h(z_h)^{3/2}  \ (z_h - z_t)   } ,     \\
T(\ta_1,\vec{x} ;  \ta_2 ,\vec{x}) &=&  \int_\e^{z_t} dz \ \fr{2 R z_t } {z \sqrt{\bar{F} z_t^2 - \bar{F}_t z^2}} \approx \fr{2 R}{  \sqrt{h(z_h)} } \log \fr{2 z_h^2}{(z_h - z_t) \e}  .   
\ee
After the Wick rotation, these relations imply the following late-time temporal correlator 
{
\be
\lim_{| t_1 - t_2 | \to \infty} G(t_1-t_2,\vec{0})  \sim  \fr{1}{| t_1 - t_2 |^{2 \D_{\rm IR} } } ,     \la{Result:Autocorrinext}
\ee
where the leading large-$\D$ geodesic exponent is the same $\D_{\rm IR}$ defined in Eq.~\eq{Throat:thermalcorr}, evaluated at the extremal horizon. Here $\D_{\rm IR}$ is the leading geodesic exponent, while the exact IR conformal dimension $\d_{\rm IR}=1/2+\sqrt{1/4+m^2R_2^2}$ was defined in the near-extremal throat analysis. In the heavy-operator limit, $\d_{\rm IR}=\D_{\rm IR}+O(\D^0)$, so the geodesic result reproduces the leading large-$\D$ behavior \cite{Faulkner:2009wj,Iqbal:2009fd,Faulkner:2011tm,Arnaudo:2024sen}.}

\subsubsection{Temperature-dependent late-time behavior in RN-AdS$_4$}

We now specialize the general discussion to the planar Reissner-Nordstrom AdS$_4$ (RN-AdS$_4$) black hole. We define the dimensionless radial coordinate $y=z/z_h$, the dimensionless charge parameter $Q$, and the boundary chemical potential $\mu$. In the normalization used here, the blackening factor and thermodynamic scales are
\be
f(y)=1-(1+Q^2)y^3+Q^2y^4, \quad T_H=\fr{3-Q^2}{4\pi z_h}, \quad {\rm and}\quad \mu=\fr{Q}{z_h}. \la{Result:RNthermodynamics}
\ee
The Schwarzschild-connected saddle has the decay constant in Eq.~\eq{Result:correlation11}. Let $y_c$ denote the complex double-turning point of the full radial saddle. The matching coefficient is
\be
{\cal M}_{RN}(Q)=\fr{2}{3-Q^2}
\left[-{\rm Im}\sqrt{\fr{2}{y_c^2}-\fr{1+Q^2}{2}y_c}\right] \quad {\rm with}\quad
2Q^2y_c^4-(1+Q^2)y_c^3-2=0, \la{Result:MRNmain}
\ee
The relevant root is chosen continuously from $y_c=2^{1/3}e^{i\pi/3}$ at $Q=0$. Thus, we obtain
\be
\Gamma_{\rm SC}=2\pi T_H\D\,{\cal M}_{RN}(Q). \la{Result:GammaSCRNmain}
\ee
The derivation of ${\cal M}_{RN}$ from the full complex saddle is given in Appendix~\ref{Appendix:matching}.

For sufficiently large $T_H/\mu$, the geometry is away from extremality and no parametrically separated AdS$_2$ throat exists. Let $A_{\rm SC}$ denote the complex saddle amplitude and let $\Omega_{\rm SC}$ denote the real part of its frequency. The controlled asymptotically late-time sector considered here is then the Schwarzschild-connected saddle
\be
G(t,\vec{0})\sim A_{\rm SC}e^{-i\Omega_{\rm SC}t}e^{-\Gamma_{\rm SC}t} \quad {\rm with}\quad t=|t_1-t_2|, \la{Result:RNhighT}
\ee
Only $\Gamma_{\rm SC}$ is relevant for the decay envelope discussed below.
As $T_H/\mu$ is lowered toward extremality, an AdS$_2$ throat develops outside the Rindler cap. In the leading near-extremal limit, $h_h\to6$ for RN-AdS$_4$, and Eq.~\eq{Throat:decayconstant} gives
\be
\Gamma_{\rm throat}=\fr{2\pi T_H\D}{\sqrt6}. \la{Result:RNthroatrate}
\ee
The Schwarzschild-connected saddle and the Lorentzian AdS$_2$ throat sector therefore coexist near extremality. The existence of the throat sector is fixed by the retarded AdS$_2$ pole structure, while its matched amplitude in the full boundary correlator is determined by the outer geometry. When both sectors contribute, let $A_{\rm throat}$ denote the matched throat amplitude. The late-time correlator may then be written schematically as~\cite{Fidkowski:2003nf,Festuccia:2005pi,Festuccia:2008zx,Faulkner:2009wj,Faulkner:2011tm}
\be
G(t,\vec{0})\sim A_{\rm SC}e^{-i\Omega_{\rm SC}t}e^{-\Gamma_{\rm SC}t}
      +A_{\rm throat}e^{-\Gamma_{\rm throat}t}. \la{Result:RNtwosaddles}
\ee
The comparison below concerns only the exponential damping rates, and the relative phases and amplitudes are fixed by the full Lorentzian matching.
Near extremality, ${\cal M}_{RN}\simeq2C_\star/(3-Q^2)$ with $C_\star\simeq2.072$ in Appendix~\ref{Appendix:limits}. Using Eq.~\eq{Result:RNthermodynamics}, the Schwarzschild-connected rate approaches
\be
\Gamma_{\rm SC}\to
\fr{C_\star\D\mu}{\sqrt3} \quad {\rm for}\quad T_H/\mu\to0 \quad {\rm and}\quad Q^2\to3^- , \la{Result:GammaSCextphysical}
\ee
while $\Gamma_{\rm throat}\to0$. Hence, sufficiently close to extremality, $\Gamma_{\rm throat}<\Gamma_{\rm SC}$. If the matched throat amplitude is nonzero, the throat contribution therefore dominates the correlator in the asymptotically late-time limit $t\to\infty$, while the Schwarzschild-connected contribution remains exponentially subleading.

At $T_H=0$, the extremal AdS$_2$ saddle gives the power law derived in Eq.~\eq{Result:Autocorrinext}. This zero-temperature result should not be identified with the $T_H\to0$ continuation of the Schwarzschild-connected saddle, since the two belong to different saddle sectors. Thus the RN-AdS$_4$ temporal correlator has three regimes. At high temperature, the controlled sector considered here is the Schwarzschild-connected exponential. Near extremality, the two exponentially damped sectors can coexist, and a nonzero throat contribution dominates only at asymptotically late time. At extremality, the temporal response is governed by the AdS$_2$ power law. The quantitative separation of the near-extremal throat regime and the finite-time competition of saddles are discussed in Appendix~\ref{Appendix:limits}.

\section{{Magnetic-field-induced lower-dimensional behavior}}

We next consider a magnetic RG flow whose ultraviolet fixed point is a four-dimensional CFT. Its holographic description is five-dimensional Einstein-Maxwell theory with AdS$_5$ radius $R$. We denote the five-dimensional Newton constant by $G_5$, the Ricci scalar by $\mathcal{R}$, and the bulk gauge potential by $A_\mu$. Its field strength is $F_{\mu\nu}=\partial_\mu A_\nu-\partial_\nu A_\mu$, and the cosmological constant is $\Lambda=-6/R^2$. The action is
{
\be
S=\frac{1}{16\pi G_5}\int d^5 x\sqrt{-g}\biggl(\mathcal{R}-2\Lambda-F^{\mu\nu}F_{\mu\nu}\biggr)  ,
\ee
}
The Einstein and Maxwell equations are
\be
\mathcal{R}_{\mu\nu}-\frac{1}{2}(\mathcal{R}-2\Lambda)g_{\mu\nu}&=&2 \ls F_{\mu\rho} {F_\nu}^\rho-\frac{1}{4}g_{\mu\nu}F^{\rho\sigma}F_{\rho\sigma} \rs , \\
\color{black}\nabla_\mu F^{\mu\nu}&=&\color{black}0.
\ee
We turn on a constant magnetic field,
\be
F= B \ d x^1 \wedge d x^2 ,
\ee
which breaks the boundary Lorentz symmetry. We denote the direction parallel to the magnetic field by $y$ and the transverse coordinates by $\vec{x}=(x_1,x_2)$. In this section we therefore write the correlator as $G(\Delta t,\Delta y,\Delta\vec{x})$.

An asymptotically AdS solution of this theory is dual to a four-dimensional UV CFT. We use a radial coordinate $r$ for which $r\to\infty$ is the UV region and $r\to0$ is the IR region. A metric ansatz that supports this magnetic flow is
\be
ds^2&=&- e^{2W(r)}  dt^2+  e^{- 2 W(r)}  dr^2+ e^{2W(r)} dy^2 +e^{2V(r)} \ls  {d x_1}^2+{d x_2}^2\rs .
\ee
For $B=0$, the gravity theory admits five-dimensional AdS space with $e^{2 V(r)} = e^{2W(r)} = r^2/R^2 $. {For nonzero $B$, the evolution of the metric in the radial direction is governed by
\be
V'' &=& e^{-2W}\left(\fr{4}{R^2}-\fr{4}{3} B^2 e^{-4V}\right)-2V'^2-3V'W', \nn
W'' &=& e^{-2W}\left(\fr{4}{R^2}+\fr{2}{3} B^2 e^{-4V}\right)-3W'^2-2V'W' ,  \la{Equation:zerotemp}
\ee
where the prime denotes a derivative with respect to $r$.  These equations must be supplemented by the first-order Einstein constraint
\be
V'^2+W'^2+4V'W'
-e^{-2W}\left(\fr{6}{R^2}-B^2e^{-4V}\right)=0 .
\la{Equation:constraintmag}
\ee
Eq.~\eq{Equation:zerotemp} together with the constraint in Eq.~\eq{Equation:constraintmag} reproduces the standard zero-temperature magnetic-brane equations of Refs.~\cite{DHoker:2009mmn,DHoker:2009ixq}.} This system admits a solution that interpolates from AdS$_5$ in the UV to AdS$_3 \times \mathbb{R}^2$ in the IR. In the asymptotic region $r\to\infty$, the perturbative solution contains integration constants $v$ and $c_1$ and takes the form
\be
e^{2 V} &=&   B v \fr{r^2}{R^2} \lb 1  + \fr{1}{r^4} \ls c_1+\fr{R^6}{3v^2}\log \fr{r}{R} \rs + \cdots \rb , \nn
e^{2 W} &=&  \fr{r^2}{R^2} \lb 1  -   \fr{1}{r^4} \ls 2c_1 + \fr{2R^6}{3v^2}\log \fr{r}{R} \rs + \cdots \rb  ,  \la{Result:UVmetric}
\ee
where the ellipses denote higher-order corrections. The asymptotic AdS normalization requires $v=1/B$, for which the metric approaches 
\be
ds^2&=& -  \fr{r^2}{R^2} dt^2+ \fr{R^2} {r^2} dr^2 + \fr{r^2}{R^2} dy^2 + \fr{r^2}{R^2} \ls  {d x_1}^2+{d x_2}^2\rs  ,    \la{Metric:asymptotic}
\ee
which is AdS$_5$. 

\begin{figure}
	\begin{center}
		\vspace{-1cm}
		\hspace{-0.cm}
		\subfigure[$V(r)$ and $W(r)$]{\label{fig1a} \includegraphics[angle=0,width=0.47\textwidth]{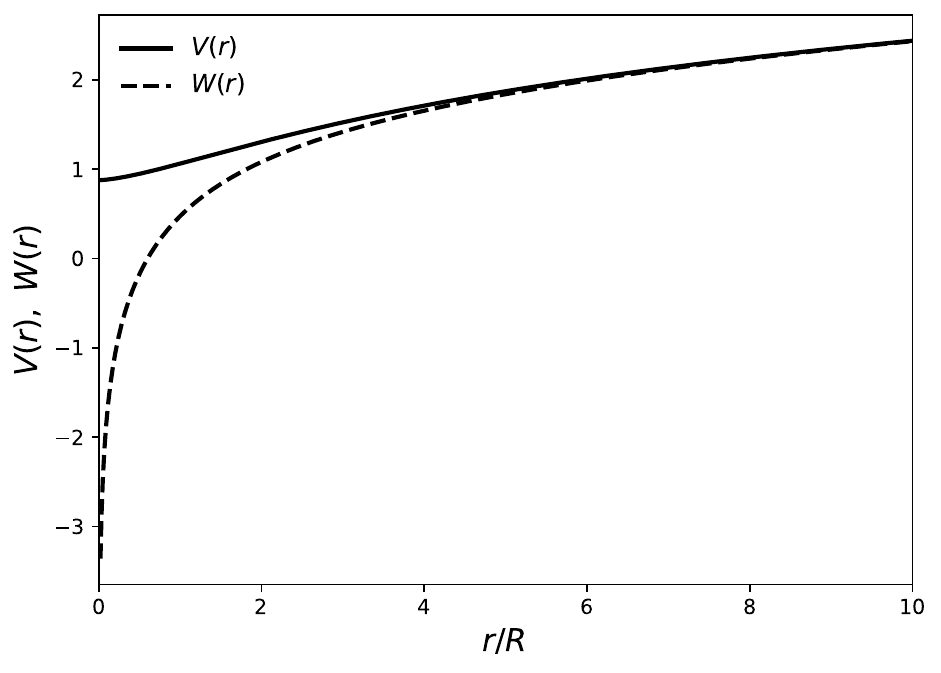}}
		\hspace{0.2cm}
		\subfigure[$d e^V/d(r/R)$ and $d e^W/d(r/R)$]{\label{fig1b} \includegraphics[angle=0,width=0.47\textwidth]{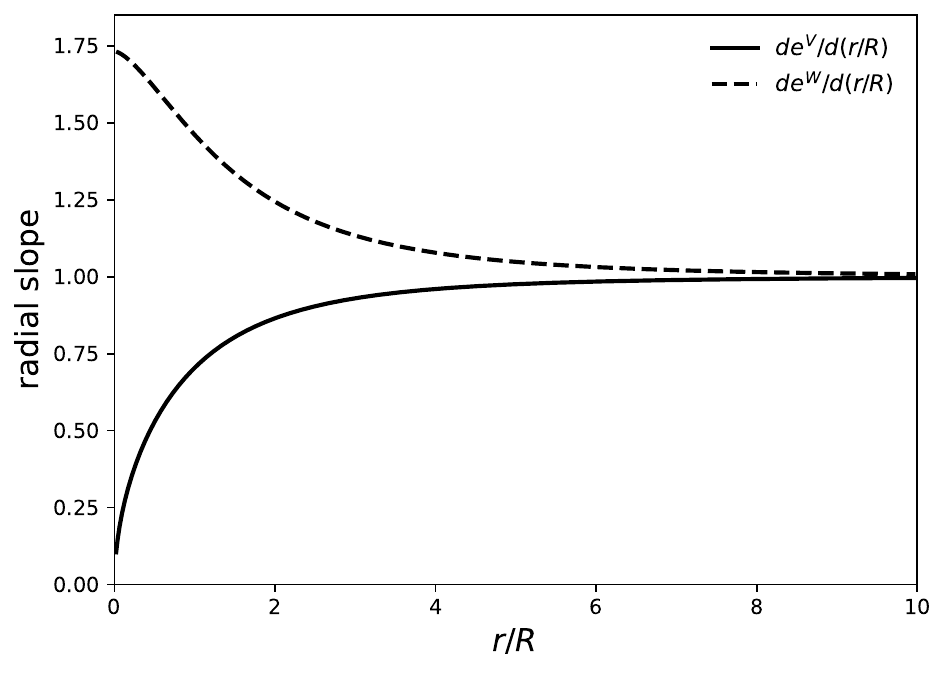}} 
		\vspace{0cm}
			\caption{\small {Numerical magnetic-brane flow for $R=1$, $B=10$, and $c_2\approx0.4344$. Panel (a) shows $V(r)$ and $W(r)$ as functions of $r/R$. Panel (b) shows $d e^V/d(r/R)$ and $d e^W/d(r/R)$. The two slopes approach one in the UV, consistent with the asymptotic AdS$_5$ normalization.}}
		\label{fig1}
	\end{center}
\end{figure}

\begin{figure}
	\begin{center}
		\vspace{-1cm}
		\hspace{-1.cm}
		\subfigure[Normalized suppression power]{\label{fig2a} \includegraphics[angle=0,width=0.47\textwidth]{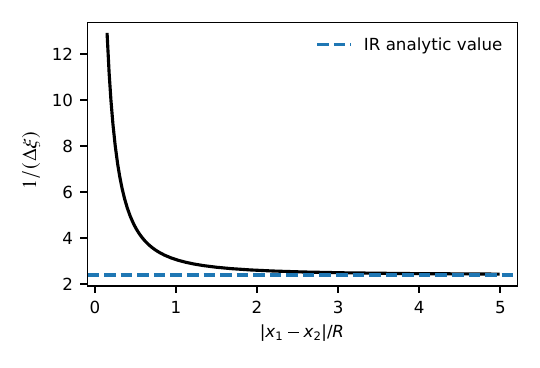}}
		\hspace{0.15cm}
		\subfigure[Normalized effective exponent]{\label{fig2b} 
		\begin{overpic}[angle=0,width=0.47\textwidth]{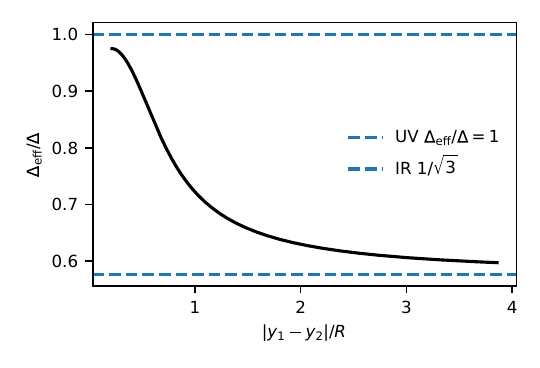}
		\end{overpic}} 
		\vspace{0cm}
		\caption{\small {Leading large-$\Delta$ geodesic quantities. (a) shows the transverse normalized screening rate $1/(\Delta \xi)$. (b) shows the longitudinal normalized effective exponent $\Delta_{\rm eff}/\Delta$. The dashed IR values are $\sqrt{B}/(3^{1/4}\sqrt R)$ and $1/\sqrt3$.}}
		\label{fig2}
	\end{center}
\end{figure}

In the IR region $r\to0$, we parameterize the leading irrelevant deformation by its amplitude $c_2$, its radial exponent $\sigma$, and the relative deformation coefficient $\kappa$. Linearizing the equations around the AdS$_3\times \mathbb{R}^2$ fixed point gives
\be
e^{2 V} &=& \fr{B R}{\sqrt{3}}  \ls 1 + c_2 r^{\sigma} + \cdots \rs , \nn
e^{2 W} &=& 3 \fr{r^2}{R^2} \ls 1 + \kappa c_2 r^{\sigma} + \cdots \rs ,   \la{Result:IRmetric}
\ee
The exponent $\sigma$ controls the leading radial departure from the fixed point, and $\kappa$ fixes the relative deformation of $W$ with respect to $V$. Substitution of the linearized mode into the equations and the constraint gives
\be
3\sigma^2+6\sigma-16=0 \quad {\rm with}\quad
\kappa=\fr{2(1-\sigma)}{1+\sigma},
\ee
so that the regular irrelevant branch is
\be
\sigma=\fr{\sqrt{57}-3}{3}>0 \quad {\rm and}\quad
\kappa=-2+\fr{4\sqrt{57}}{19}.
\ee
The amplitude $c_2$ is fixed by matching the regular IR mode to the UV solution. The AdS$_3$ factor has infrared radius
\be
R_{IR} = \fr{R}{\sqrt{3}} .
\ee
Using this radius, the fixed-point geometry is
\be
ds^2 = -  \fr{r^2}{R_{IR}^2} dt^2+ \fr{R_{IR}^2} {r^2} dr^2 + \fr{r^2}{R_{IR}^2} dy^2  +  B R_{IR}  \ls  {d x_1}^2+{d x_2}^2\rs  .     \la{Metric:IRregion}
\ee
The interpolating solution used in Figs.~\ref{fig1} and \ref{fig2} is obtained with $R=1$, $B=10$, and an initial radius $r_0/R=10^{-4}$. We integrate outward from the IR expansion in Eq.~\eq{Result:IRmetric} to a large-radius UV endpoint $r_{\max}$ and impose $V(r_{\max})-W(r_{\max})=0$. This fixes $c_2\approx0.4344$. These parameters define the numerical illustration used below.

We now probe this full flow with a minimally coupled neutral scalar of bulk mass $m$. Its UV conformal dimension $\D$ is defined by the AdS$_5$ mass-dimension relation $m^2R^2=\D(\D-4)$. In the heavy-operator limit considered here, $\D\gg1$. In the UV, where the geometry approaches AdS$_5$, the two-point function reduces to the CFT form 
\be
G(0,\Delta y,\Delta\vec{x})  &\sim& \fr{1}{ \ls  |\vec{x}_1 - \vec{x}_2|^2 + |y_1 - y_2|^2 \rs^{\D} }   .
\ee
The effect of the magnetic field is negligible in this UV limit. In the IR limit, however, the transverse and longitudinal correlators behave differently because rotational symmetry is broken, as shown in Eq.~\eq{Metric:IRregion}. We first consider the transverse correlator in the $\vec{x}$ direction. We write $U(r)\equiv e^{2W(r)}$ and choose the separation along one transverse coordinate $x$. The corresponding geodesic curve lies in the $r$-$x$ plane and has length
\be
L (y,\vec{x}_1;y,\vec{x}_2) = \int_{x_1}^{x_2} dx \sqrt{ \fr{r'^2}{U} + e^{2 V}} ,    \la{GeodesicL:x}
\ee
where the prime denotes a derivative with respect to $x$. Using Eq.~\eq{Result:IRmetric}, the leading contribution in the IR region ($r \to 0$) is approximated by
{
\be
L (y,\vec{x}_1;y,\vec{x}_2) \approx \int_{x_1}^{x_2} dx \sqrt{ \fr{R^2 r'^2}{3r^2\left(1+\kappa c_2r^\sigma\right)} +  \fr{B R}{\sqrt{3}}  \left(1+c_2r^\sigma\right) } .  \la{GeodesicL:xIR}
\ee
}

We restrict the IR radial region to $0<r\le r_m$ and take the turning point to satisfy $r_t<r_m$. The real radial integration range is then $r_t\le r\le r_m$. Using the conserved quantity, the boundary separation and geodesic length become
{
\be
| \vec{x}_1 - \vec{x}_2 |  &=& \int_{r_t}^{r_m} dr\, \fr{2\sqrt{R}\sqrt{1+c_2r_t^\sigma}}{3^{1/4}\sqrt{B}\sqrt{c_2}\,r\sqrt{1+\kappa c_2r^\sigma}\sqrt{1+c_2r^\sigma}\sqrt{r^\sigma-r_t^\sigma}}+\cdots, \nn
L (y,\vec{x}_1;y,\vec{x}_2) &=& \int_{r_t}^{r_m} dr\, \fr{2R\sqrt{1+c_2r^\sigma}}{\sqrt{3}\sqrt{c_2}\,r\sqrt{1+\kappa c_2r^\sigma}\sqrt{r^\sigma-r_t^\sigma}}+\cdots .
\ee
}
The long-distance IR limit corresponds to $r_t\to0$. Evaluating the leading terms gives
\be
G(0,0,\Delta\vec{x}) \sim e^{-| \vec{x}_1 - \vec{x}_2 | / \xi  }  ,
\ee
where the correlation length is given by
\be
\fr{1}{\xi}  = \fr{\sqrt{B} \, \D }{ 3^{1/4} \sqrt{R}}  .    \la{Result:cfun}
\ee
{Since $\sigma>0$, the irrelevant mode gives only subleading corrections proportional to $r_t^\sigma$. Therefore, the leading exponential transverse IR behavior is fixed by the AdS$_3\times \mathbb{R}^2$ fixed-point geometry.}
{Because the geodesic approximation is controlled only for $\D\gg1$, Fig.~\ref{fig2}(a) shows the normalized large-$\D$ quantity
\be
\fr{1}{\D\,\xi}=-\fr{1}{\D}\fr{\pa \log G(0,0,\Delta\vec{x})}{\pa |\vec{x}_1-\vec{x}_2|} .
\ee
For the parameters used in the numerical illustration, $R=1$ and $B=10$, the analytic IR value is $1/(\D\xi)=\sqrt{B}/3^{1/4}\simeq2.403$, in agreement with Fig.~\ref{fig2}(a).}

We next turn to the longitudinal IR correlator,   
\be
L (y_1,\vec{x};y_2,\vec{x}) = \int_{y_1}^{y_2} dy \sqrt{\fr{r'^2}{U}+ e^{2 W}} ,    \la{GeodesicL:y}
\ee
where the prime denotes a derivative with respect to $y$. In the IR, the leading contribution is governed by the AdS$_3$ part of Eq.~\eq{Metric:IRregion}. Using the translational symmetry in the $y$-direction with $R_{IR}=R/\sqrt{3}$, the separation and length are

\be
|y_1-y_2| &=& 2\int_{r_t}^{r_m} dr \,
\fr{R_{IR}^2 r_t}{r^2\sqrt{r^2-r_t^2}}
=\fr{2R_{IR}^2}{r_t}+O(r_t/r_m^2), \nn
L (y_1,\vec{x};y_2,\vec{x}) 
&=& 2R_{IR}\int_{r_t}^{r_m} \fr{dr}{\sqrt{r^2-r_t^2}}
=2R_{IR}\log \fr{2r_m}{r_t}+\cdots .
\ee
Eliminating $r_t$ in favor of $|y_1-y_2|$ and introducing a finite matching scale $\L_m$ associated with the radial matching point $r_m$, we obtain
\be
L (y_1,\vec{x};y_2,\vec{x}) \approx \fr{2 R}{\sqrt{3}} \log \ls 3 \L_m |y_1 - y_2 | \rs  + \cdots . 
\ee
This expression applies in the long-distance limit $|y_1-y_2|\to\infty$. The leading large-$\D$ geodesic exponent is $\D_{\rm IR}^{\rm geo}=\D/\sqrt{3}$. After subtracting the UV divergence, the longitudinal correlator becomes
\be
G(0,\Delta y,\vec{0}) \sim \fr{1}{   |y_1 - y_2| ^{2 \D_{\rm IR}^{\rm geo}} } . 
\ee As in Sec.~2, this is distinct from the exact IR conformal dimension $\d_{\rm IR}$. For the minimally coupled neutral probe considered here, the bulk mass $m$ is unchanged along the flow. The scalar wave equation in AdS$_3$ therefore gives
\be
\d_{\rm IR}(\d_{\rm IR}-2)=m^2 R_{IR}^2 .
\ee
Combining this AdS$_3$ relation with the UV definition $m^2R^2=\D(\D-4)$ and using $R_{IR}=R/\sqrt{3}$, the exact IR dimension is $\d_{\rm IR}=1+\sqrt{1+\D(\D-4)/3}=\D/\sqrt{3}+O(\D^0)$. For $\D \gg 1$, therefore, $\d_{\rm IR}=\D_{\rm IR}^{\rm geo}+O(\D^0)$, confirming the geodesic result at leading order in the heavy-operator limit. The zero-temperature magnetic background preserves boost symmetry in the $(t,y)$ plane, while the AdS$_3$ factor makes the corresponding $1+1$-dimensional conformal structure manifest in the IR. The general IR two-point function is 
\be
G(\Delta t,\Delta y,\vec{0}) \sim \fr{1}{ \ls - | \Delta t-i\epsilon |^2 + |\Delta y|^2 \rs^{\d_{\rm IR}} } ,
\ee
which has the two-dimensional CFT form. Thus, the transverse correlator is exponentially suppressed, whereas the longitudinal correlator exhibits CFT$_2$ scaling with the infrared conformal dimension $\d_{\rm IR}$. 

To check this anisotropic long-distance behavior, we evaluate the longitudinal geodesic numerically using Eq.~\eq{GeodesicL:y} and then determine an effective conformal dimension \cite{Park:2022mxj,Park:2024fik}
\be
\D_{eff}= -  \fr{1}{2} \fr{\pa \log G(0,\Delta y,\vec{0})}{\pa \log |y_1 -y_2| }  .
\ee
We use the same numerical procedure as in the transverse case. Fig.~\ref{fig2}(b) shows leading large-$\D$ geodesic quantities. In the IR limit, $\D_{eff}/\D$ approaches $1/\sqrt3$. Thus, the longitudinal heavy-scalar correlator approaches the infrared CFT$_2$ scaling form.


\section{Discussion}
 
In this work, we have shown that a factorized lower-dimensional infrared geometry does not impose a single universal infrared scaling law on all kinematic sectors of a heavy-scalar two-point function. Instead, the infrared behavior depends on temperature, spacetime direction, and the relevant bulk saddle. Importantly, this does not imply a dimensional reduction of the microscopic QFT. Rather, a lower-dimensional conformal theory can emerge as an effective infrared description of selected sectors while the underlying theory remains defined in its original spacetime dimension.

For charged asymptotically AdS black holes, the equal-time spatial correlator remains exponentially screened even at extremality. In contrast, the temporal correlator exhibits different late-time regimes as the temperature is lowered. The Schwarzschild-connected saddle has a decay rate that requires matching to the full geometry, while near extremality an AdS\(_2\) throat supports an additional sector whose leading decay rate is fixed by the throat data. In the heavy-operator limit, this rate agrees with the characteristic lowest damping scale of the isolated finite-temperature AdS\(_2\) Green function. For RN-AdS$_4$, the throat contribution becomes dominant at asymptotically late times when its matched amplitude is nonzero. At extremality, the thermal exponential is replaced by the AdS$_2$ power law. The AdS\(_2\) throat therefore provides an effective one-dimensional conformal sector in the infrared temporal response.

The magnetic-brane flow gives a more direct realization of an effective lower-dimensional CFT. The geometry interpolates from AdS\(_5\) in the ultraviolet to AdS\(_3\times\mathbb{R}^2\) in the infrared. Using the same neutral heavy-scalar probe in both directions, the transverse correlator remains exponentially screened, whereas the longitudinal correlator approaches the power law dictated by the AdS\(_3\) factor. Thus, an effective CFT\(_2\) emerges in the infrared of a microscopic four-dimensional QFT. The lower-dimensional scaling is associated with the AdS\(_3\) sector, while the transverse directions remain tied to the spectator \(\mathbb{R}^2\) factor.

Taken together, these examples show that factorized infrared geometries can support effective lower-dimensional conformal dynamics without microscopic dimensional reduction. Which scaling behavior is realized is determined by the kinematic sector, temperature, and saddle structure. In this sense, holographic RG flows provide a geometric mechanism through which lower-dimensional CFT behavior can emerge dynamically in the infrared of a higher-dimensional quantum field theory.


\section*{Acknowledgments}

This work was supported by the National Research Foundation of Korea (NRF) grant funded by the Korea government (MSIT) (No. RS-2026-25473667).

\appendix
\section{Finite matching from full complex saddles}\la{Appendix:matching}

We now determine the full-geometry matching coefficient ${\cal M}$ introduced in Sec.~2 from the corresponding complex Lorentzian saddle. In the large-$\Delta$ geodesic/WKB limit at zero spatial momentum, a complex saddle with critical energy $E_c$ contributes to the late-time correlator. Its decay rate is fixed by the imaginary part of $E_c$. Comparing this rate with the Schwarzschild-connected decay law in Eq.~\eq{Result:correlation11} gives
\be
G(t)&\sim&e^{-iE_ct} \quad {\rm with} \quad
\Gamma_{\rm SC} = -{\rm Im}E_c, 
\ee
where the decay constant $\Gamma_{\rm SC}$ is related to ${\cal M}$ in Eq.~\eq{Result:correlation11}, which measures the finite full-geometry correction,
\be
{\cal M}&=&\fr{\Gamma_{\rm SC}}{2\pi T_H\Delta}
=-\fr{{\rm Im}E_c}{2\pi T_H\Delta}. \la{Appendix:Mdef}
\ee
It is not fixed by the near-horizon Rindler expansion alone.

To determine $E_c$, consider Schwarzschild gauge with $ds^2=-F(r)dt^2+dr^2/F(r)+\cdots$. For a massive probe at zero spatial momentum, the radial Hamilton-Jacobi equation has turning points satisfying $E^2=m^2F(r)$. For a fixed complex energy $E$, this equation is understood in the complexified radial plane and generally has several radial roots for the black-brane geometries considered here. The two turning points relevant to a given WKB saddle are a pair of distinct radial roots $r_1(E)$ and $r_2(E)$. In the large-time limit, the boundary time integral becomes singular when the relevant pair of radial turning points coalesces. As $E$ approaches the critical value $E_c$, one has $r_1(E),r_2(E)\to r_c$. The corresponding double turning point therefore determines the critical complex energy that controls the asymptotic exponential decay. Hence
\be
E_c^2=m^2F(r_c) \quad {\rm and}\quad F'(r_c)=0,
\ee
with $mR\simeq\Delta$. In other coordinates, the same condition is expressed as a double turning point of the corresponding radial effective potential.

For the planar Schwarzschild-AdS$_{d+1}$ black brane with horizon radius $r_h$,
\be
F(r)=\fr{r^2}{R^2}\lb1-\ls\fr{r_h}{r}\rs^d\rb \quad {\rm with}\quad
T_H=\fr{d r_h}{4\pi R^2},
\ee
the turning-point equation can be written as
\be
r^d-\fr{E^2R^2}{m^2}r^{d-2}-r_h^d=0.
\ee
Thus the turning points are naturally roots in the complex $r$-plane. For generic complex $E$, this degree-$d$ equation has several roots, and the late-time saddle selects the pair that coalesces at $E=E_c$. A massive zero-momentum particle with $S=-Et+S_r(r)$ then obeys the double-turning-point condition above. The retarded branch is
\be
r_c=r_h\ls\fr{d-2}{2}\rs^{1/d}e^{-i\pi/d} \quad {\rm and}\quad
E_c=\fr{m r_h}{R}\ls\fr{d-2}{2}\rs^{1/d}\sqrt{\fr{d}{d-2}}e^{-i\pi/d} .
\ee
Using Eq.~\eq{Appendix:Mdef}, the decay constant can be written as $\Gamma_{\rm SC}^{(d)}=2\pi T_H\D\,{\cal M}_d$ with
\be
{\cal M}_d=\fr{2}{d}\ls\fr{d-2}{2}\rs^{1/d}\sqrt{\fr{d}{d-2}}\sin\fr{\pi}{d} \quad {\rm for}\quad d>2,
\ee
where $mR\simeq\D$. This reproduces the large-mass WKB result \cite{Fidkowski:2003nf,Festuccia:2005pi,Festuccia:2008zx}. The $d\to2$ limit gives ${\cal M}_2=1$ and $d=4$ gives ${\cal M}_4=1/2$.

For the planar RN-AdS$_4$ black brane, write
\be
f(y)=1-(1+Q^2)y^3+Q^2y^4 \quad {\rm with}\quad
y=\fr{z}{z_h} \quad {\rm and}\quad T_H=\fr{3-Q^2}{4\pi z_h} . \la{Appendix:RNf}
\ee
For a neutral heavy scalar at zero spatial momentum, the Schwarzschild-connected complex saddle is fixed by the double turning point of $f(z)/z^2$,
\be
2Q^2y_c^4-(1+Q^2)y_c^3-2=0, \la{Appendix:RNroot}
\ee
with the root chosen continuously from $y_c=2^{1/3}e^{i\pi/3}$ at $Q=0$. The corresponding decay constant is
\be
\Gamma_{\rm SC}=2\pi T_H\D\,{\cal M}_{RN}(Q) \quad {\rm with}\quad
{\cal M}_{RN}(Q)=\fr{2}{3-Q^2}\left[-{\rm Im}\sqrt{\fr{2}{y_c^2}-\fr{1+Q^2}{2}y_c}\right] .
\ee
Therefore, the charge changes the finite matching coefficient but not the form of the non-extremal decay law on this branch. At $Q=0$, ${\cal M}_{RN}=2^{-1/3}$, while $Q^2=1$ gives ${\cal M}_{RN}\simeq1.612$.

The near-extremal analysis clarifies the relation to the zero-temperature geometry. Along the same Schwarzschild-connected branch,
\be
Q^2\to3^- \quad {\rm with}\quad
{\cal M}_{RN}\simeq\fr{2C_\star}{3-Q^2} \quad {\rm and}\quad
\Gamma_{\rm SC}\to\fr{\D}{z_h}C_\star \quad {\rm where}\quad
C_\star\simeq2.072 .
\ee
Therefore, this exponential saddle keeps a finite decay constant as $T_H\to0$. At extremality,
\be
f(y)=(1-y)^2(1+2y+3y^2),
\ee
so the horizon develops the AdS$_2$ double zero with $h(z_h)=6$. The late-time correlator is then controlled by the extremal AdS$_2$ branch described in Sec.~2, with the leading geodesic exponent $\D_{\rm IR}=\D/\sqrt6$ and the exact neutral zero-momentum scalar dimension $\d_{\rm IR}=1/2+\sqrt{1/4+m^2R^2/6}$. The finite-temperature Schwarzschild-connected saddle and the extremal AdS$_2$ saddle therefore have different limiting behaviors. Appendix~\ref{Appendix:limits} shows that the near-extremal AdS$_2$ throat also supports a finite-temperature geodesic saddle with a decay rate proportional to $T_H$. Determining which branch dominates requires the global complex-saddle structure. A Stokes transition is a natural WKB mechanism for a change of dominance \cite{Edalati:2010hk,Ficek:2023qsr}.

\section{Lorentzian saddle competition and extremal limits}\la{Appendix:limits}

Section~2 obtains the Schwarzschild-connected and AdS$_2$-throat sectors by analytic continuation of Euclidean geodesics. Their Lorentzian interpretation follows from the retarded prescription. Writing $G_R$ for the retarded Green function, for $t>0$ we have
\be
G_R(t)=\int_{-\infty+i0^+}^{\infty+i0^+}\fr{d\omega}{2\pi}\,e^{-i\omega t}G_R(\omega),
\ee
and the contour closes in the lower half-plane. Quasinormal poles with nonzero residues therefore contribute directly to the late-time correlator \cite{HorowitzHubeny:1999,SonStarinets:2002,HerzogSon:2002,SkenderisVanRees:2008,SkenderisVanRees:2009,Barnes:2010}. The Schwarzschild-connected WKB saddle is continuously connected to the standard Schwarzschild-AdS retarded quasinormal branch \cite{Fidkowski:2003nf,Festuccia:2005pi,Festuccia:2008zx}. The near-extremal throat sector follows from the ingoing AdS$_2$ solution. For a neutral scalar at zero spatial momentum, we define $\nu=\sqrt{1/4+m^2R_2^2}$ and $\d_{\rm IR}=1/2+\nu$. The frequency-dependent part of the finite-temperature retarded Green function then has the standard ratio form \cite{Faulkner:2009wj,Faulkner:2011tm,Iqbal:2009fd}
\be
{\cal G}_{R}^{(2)}(\omega)\propto
(2\pi T_H)^{2\nu}
\fr{\Gamma\!\left(\d_{\rm IR}-\fr{i\omega}{2\pi T_H}\right)}
{\Gamma\!\left(1-\d_{\rm IR}-\fr{i\omega}{2\pi T_H}\right)} ,
\ee
up to factors independent of $\omega$. For generic $\nu$, the numerator gives the isolated AdS$_2$ poles $\omega_n^{(2)}=-i\,2\pi T_H(n+\d_{\rm IR})$ with $n=0,1,2,\ldots$. These poles determine the characteristic thermal scales of the throat Green function. Whether a corresponding pole contributes to the full boundary correlator with nonzero residue depends on the matching to the outer region. When the matched throat sector is present, its lowest infrared damping scale is $2\pi T_H\d_{\rm IR}$. Since $\d_{\rm IR}=\D_{\rm IR}+O(\Delta^0)$ in the heavy-operator limit, this damping scale agrees with the analytically continued geodesic result at late times to leading order in large $\Delta$. The full matched residue and the precise Stokes competition with the Schwarzschild-connected branch remain global properties of the complete geometry.

\subsection{Extremal and near-extremal limits}

For RN-AdS$_4$, let $\epsilon_T=3-Q^2$ so that $T_H=\epsilon_T/(4\pi z_h)$. The exact extremal saddle gives
\be
G_0(t)\sim |t|^{-2\Delta/\sqrt6} \quad {\rm with}\quad \gamma_{\rm eff}(0,t)\equiv-\partial_t\log|G_0(t)|\to0.
\ee
At small but nonzero temperature, the AdS$_2$-throat result of Sec.~2 gives
\be
G_{\rm throat}(t)\sim
\left[\fr{\pi T_H\ell_{\rm IR}}{\sinh(\pi T_H|t|)}\right]^{2\Delta/\sqrt6} \quad {\rm with}\quad
\Gamma_{\rm throat}=\fr{2\pi T_H\Delta}{\sqrt6}\to0.
\ee
Hence the throat saddle interpolates smoothly between an intermediate-time AdS$_2$ power law and a thermal exponential. By contrast, along the Schwarzschild-connected branch of Appendix~\ref{Appendix:matching},
\be
\Gamma_{\rm SC}\to\fr{C_\star\Delta\mu}{\sqrt3} \quad {\rm with}\quad C_\star\simeq2.072,
\ee
as $T_H/\mu\to0$ and $Q^2\to3^-$. Here we used $\mu=Q/z_h$. The finite value is therefore a property of that continued saddle, not automatically the decay rate of the complete correlator. The apparent noncommutativity of $T_H\to0$ and $t\to\infty$ reflects the nonuniform behavior of distinct saddle sectors in the two limits.

\subsection{Temperature range and dominant late-time sector}

The throat is controlled only when a quadratic AdS$_2$ region is parametrically separated from the ordinary Rindler region. With $x=(z_h-z)/z_h$, the exact RN-AdS$_4$ horizon expansion is
\be
f(x)=(3-Q^2)x+3(Q^2-1)x^2+(1-3Q^2)x^3+Q^2x^4.
\ee
For $Q^2>1$, comparison of the first two terms defines
\be
x_{\rm cross}=\fr{3-Q^2}{3(Q^2-1)}.
\ee
A controlled throat requires $x_{\rm cross}\ll x\ll1$. Thus $Q^2\simeq3$ is the near-extremal throat regime. For illustration, choosing $x_{\rm cross}=0.1$ gives $Q^2\simeq2.54$, while $Q^2\leq3/2$ has no quadratic-dominated interval inside $x<1$. These values are only scale markers and do not define phase boundaries. In the normalization $\mu=Q/z_h$,
\be
\fr{T_H}{\mu}=\fr{3-Q^2}{4\pi Q},
\ee
so $Q^2\gtrsim2.54$ corresponds roughly to $T_H/\mu\lesssim2.3\times10^{-2}$ and $Q^2=3/2$ to $T_H/\mu\simeq9.7\times10^{-2}$.

Accordingly, sufficiently near extremality and for nonzero throat residue, $\Gamma_{\rm throat}\ll\Gamma_{\rm SC}$ and the throat pole controls the asymptotic decay. At intermediate $Q$ the full spectrum and residues must be compared, while toward the Schwarzschild limit no parametrically separated AdS$_2$ throat sector is controlled and the full non-extremal quasinormal spectrum governs the relaxation. A precise exchange temperature requires the full RN-AdS$_4$ quasinormal spectrum, residues, and Stokes multipliers and is not determined by the local expansion alone.

\end{document}